# Influence of encapsulation temperature on Ge:P *δ*-doped layers

G. Scappucci,[1] G. Capellini,[2] and M. Y. Simmons[1]

[1]*School of Physics and Australian Research Council Centre of Excellence for Quantum Computer Technology, University of New South Wales, Sydney, NSW 2052, Australia.*

[2]*Dipartimento di Fisica, Università di Roma Tre, Via della Vasca Navale 84, 00146 Roma, Italy.*

**Abstract**

We present a systematic study of the influence of the encapsulation temperature on dopant confinement and electrical properties of Ge:P *δ*-doped layers. For increasing growth temperature we observe an enhancement of the electrical properties accompanied by an increased segregation of the phosphorous donors, resulting in a slight broadening of the *δ*-layer. We demonstrate that a step-flow growth achieved at ~ 530 °C provides the best compromise between high crystal quality and minimal dopant redistribution, with an electron mobility ~ 128 cm$^2$/Vs at a carrier density $1.3 \times 10^{14}$ cm$^{-2}$, and a 4.2 K phase coherence length of ~ 180 nm.

---

[1] Electronic mail: giordano.scappucci@unsw.edu.au; michelle.simmons@unsw.edu.au

The recent demonstration of phosphorus in germanium (Ge:P) δ-doped layers[1] and of atomic-scale scanning tunneling microscope (STM) hydrogen lithography on Ge(001)[2] has opened the possibility of an ultra-high vacuum (UHV) STM approach to the fabrication of atomic-scale devices in Ge analogous to the one developed for atomic-scale devices in silicon.[3] This approach consists of using an STM to create a laterally patterned dopant δ-layer which is then encapsulated under a homoepitaxial layer deposited by molecular beam epitaxy. The spatially defined δ-layer is obtained by selective adsorption and incorporation of dopants on depassivated areas of a hydrogen-terminated Ge(001) surface in which H atoms have been locally removed by an STM tip. Ge-based atomic scale-devices are particularly interesting since the electronic device miniaturization to and beyond the 16-nm-node foresees the replacement of Si channel in transistors with higher mobility materials[4,5] and requires, ultimately, the development of technologies capable of miniaturizing devices at the atomic-scale.[6]

Any design of future atomic-scale Ge devices - such as ultra-shallow abrupt junctions, ballistic transistors or quantum coherent devices - requires a detailed knowledge of the carrier transport properties in the starting two-dimensional (2D) Ge:P δ-doped layer as a function of the fabrication process parameters. Previous studies on Si:P δ-doped layers have highlighted the crucial role of the temperature at which the encapsulation layer is grown ($T_g$) in determining the spatial confinement and electronic transport properties of dopants due to the interplay between epi-layer crystal quality and dopant segregation.[7,8] Ideally one would need to carefully control the process thermal budget to find the maximum temperature that guarantees encapsulation of dopants in a high quality crystal whilst minimizing dopant redistribution. In silicon this has proven to be a difficult task: Si:P δ-layers are typically encapsulated at low temperatures (~250 °C) to avoid P segregation at the expense of a rougher surface.[7-9] In this paper, instead, we demonstrate that for Ge:P δ-doped layers a compromise between high crystal quality and minimal dopant redistribution is possible at a growth temperature ~530 °C, sufficient to achieve a high-quality step-flow Ge growth. To this end we present a systematic study of the effect of Ge encapsulation temperature on the dopant confinement and electronic transport properties of Ge:P $\delta$-doped layers, where the sheet of P dopants was encapsulated by MBE at temperatures ranging from 210 °C to 600 °C. We used STM to characterize the surface morphology after encapsulation, Secondary Ion Mass Spectroscopy (SIMS) to analyze the dopant spatial localization, and magnetotransport at 4.2 K to investigate the electrical properties of the $\delta$-layers. SIMS analysis was carried out with a $Cs^+$ primary ion beam at a low energy of 1 keV to optimize depth resolution. The magnetotransport measurements at 4.2 K were performed by means of trench-isolated Hall bars with Al ohmic contacts.

A clean, atomically flat Ge(001) surface was prepared by *in-situ* flash-anneal of a Ge sample passivated beforehand with a $GeO_x$ layer chemically grown *ex-situ*. Details of the UHV system, initial Ge surface preparation and STM characterization can be found elsewhere.[1,2] After saturation-dosing of the clean surface with $PH_3$ at a temperature ~100 °C, a first thermal anneal to 350 °C incorporates an initial amount of P atoms from the $PH_x$ species on the surface into the Ge substrate. The incorporated P is chemisorbed at the surface thereby minimizing segregation during the subsequent encapsulation step. After transfer under UHV to the MBE growth chamber, the sample was heated to the targeted encapsulation temperature $T_g$ at a rate of ~ 50 °C/min. and kept at $T_g$ for the following 3 min. to stabilize the temperature before growth. As a result, the surface undergoes a second thermal anneal just before encapsulation: we will discuss the critical effect of this pre-encapsulation anneal on P incorporation later. Finally, the $\delta$-layer is encapsulated with ~ 22 nm of intrinsic Ge at a growth rate of 0.13 Å/s rate, cooled down to room temperature and transferred to the chamber for STM analysis.

In Fig. 1(a)-(d) we report STM images of the surface topography of the $\delta$-layers after encapsulation at different $T_g$ along with atomic resolution close-ups and schematic diagrams of the encapsulation process as inset in each panel. For space reasons only images at selected $T_g$ are reported although the trend is consistent. Owing to the limited adatom mobility the growth for $T_g \leq 460$ °C proceeds in an island-like mode with the overgrown surface showing small mounds as observed in previous studies of low temperature Ge homoepitaxy [Fig. 1(a), (b)].[10,11] Given its limited thickness, however the encapsulating layer remains epitaxial, as confirmed by high resolution transmission electron microscopy.[1] Increasing $T_g$ in this growth regime, increases the adatom mobility on the surface resulting in an increase in the size of the islands where the dimer rows, visible in the close-up of the same images, elongate [Fig. 1 (b)]. At 530 °C [Fig. 1 (c)] we observe a transition to step-flow growth mode, which is even more evident in the 600 °C sample [Fig. 1 (d)]. The resulting surface of the encapsulation layer is atomically flat with dimer rows forming wide terraces. The observed transition temperature of 530 °C is higher than what reported in Ref [10], where a stepped Ge(001) surface was achieved by MBE deposition at 365 °C. The main origins for this temperature discrepancy might be due to the different growth rate used (~ 10× higher in that study), to the surface quality after the *in-situ*/*ex-situ* cleaning procedure, and to the effect of the P δ-doped layer on the subsequent growth. While the latter is likely to have a minor impact, since the same surface quality has been observed in δ-doped and un-doped samples, the other two origins are presently under investigation. The atomic-scale images in the insets of Fig. 1(c) and (d) reveal the presence of bright asymmetric features that we attribute to P atoms segregating toward the surface during growth to form Ge−P heterodimers, similar to the Si−P heterodimer found in the Si:P system after thermal anneal of δ-layers above 350 °C.[12] This is direct evidence at $T_g \geq 530$ °C for segregation of individual P atoms from the δ-doped layer to the Ge surface. The density of P atoms on the surface estimated from counting the Ge−P heterodimers is, however, limited to ~0.01 ML at $T_g = 530$ °C and increases to 0.04 ML at $T_g = 600$ °C.

All the SIMS $^{31}$P depth profiles [Fig. 2(a)] show an isolated $^{31}$P peak ~22 nm below the surface due to the confinement of P atoms on the starting Ge surface. The concentration at peak maximum is in the $10^{20}$ to $10^{21}$ cm$^{-3}$ range comparable to the solid-solubility limit of $2\times 10^{20}$ cm$^{-3}$ for P in Ge.[13] Fig. 2(b) shows the full width at half maximum (FWHM) of the dopant depth profiles, together with its leading and trailing exponential edge slopes ($\lambda_L$, $\lambda_T$), i.e. the distance over which the concentration at peak maximum is reduced by 1/e. For $T_g \leq 460$ °C, the peaks show a constant resolution-limited FWHM of ~2 nm. In this temperature range, the profiles are skewed away from the surface with constant trailing slopes $\lambda_T \sim 1$ nm greater than the leading slope $\lambda_L \sim 0.6$ nm. This is a well known result of the SIMS broadening from ion beam mixing in the sputtering process.[14] In agreement with direct atomic-scale observation with the STM, SIMS analysis therefore confirms that a temperature $T_g > 460$ °C is needed to observe a change in dopant segregation towards the surface during MBE growth. At $T_g = 530$ °C the peak FWHM has broadened to 3.65 nm, greater than the 2-nm SIMS resolution quoted above. The trailing slope $\lambda_T$ is again ~1 nm - as for the lower $T_g$ - but the profile is now skewed towards the surface indicating dopant segregation with a characteristic length given by $\lambda_L = 2.45 \pm 0.06$ nm, within the ~ 5 nm Bohr radius for P donors in Ge. At 600 °C the process of segregation is enhanced with $\lambda_L = 4.03 \pm 0.08$ nm and peak FWHM = 5.5 nm. A comparison between Ge:P and Si:P $\delta$-layers fabricated with an analogous approach shows that P migration in Ge is greatly suppressed up to $T_g = 530$°C. For the Si:P system, dopant segregation towards the surface starts to be observed above growth temperature of 270 °C[8] and becomes significant at $T_g \geq$ 400 °C with $\lambda_L \sim 100$ nm and a demise of confinement of the $\delta$-layer.[7]

The total concentration $N_{SIMS}$ of P atoms in the $\delta$-layer, obtained by integrating the depth profiles in Fig. 2(a), is reported in Fig. 2(c) as a function of $T_g$ along with the *electrically active* 2D carrier concentration $n_{Hall}$ determined from 4.2 K Hall effect measurements. There is a complex trend in $N_{SIMS}$ with increasing $T_g$, which can be explained if we consider the impact of the second pre-encapsulation anneal at the growth temperature. The chemical reactions of PH$_3$ and PH$_x$ molecules on the Ge surface induced by the anneals prior to encapsulation are critical in determining the total concentration of P atoms incorporated in the $\delta$-layer. Assuming a similar mechanism of incorporation of P atoms from the PH$_3$ adsorbate into the surface to that observed in silicon,[15] the strong dependence of $N_{SIMS}$ with $T_g$ indicates that the first thermal anneal to 350 °C does not fully incorporate the P atoms from the dopant precursor into substitutional lattice sites. As $T_g$ is increased from 210 °C to 350 °C $N_{SIMS}$ decreases from $2.39 \times 10^{14}$ cm$^{-2}$ to $1.52 \times 10^{14}$ cm$^{-2}$. Parallel thermal programmed desorption studies to be published elsewhere confirm that this decrease of incorporated P is due to the desorption of PH$_3$ during the pre-encapsulation anneal. This finding is in agreement with the results from high-resolution core-level photoemission experiments performed by Tsai *et al.* on PH$_3$-saturated Ge(001).[16] Above 350 °C, hydrogen bonded to the Ge desorbs directly as H$_2$,[17] thereby increasing the density of Ge sites available for P incorporation during the pre-encapsulation thermal anneal and, as a consequence, increasing $N_{SIMS}$. Finally in the 460 °C to 600 °C temperature range this increase is counterbalanced by the increased P desorption from the surface prior to growth, leading to an overall decrease of $N_{SIMS}$ for increasing temperature similar to that observed in Si.[7]

If we consider the electrically-active carrier concentration $n_{Hall}$, we do not observe a strong dependence with $T_g$ in the 210 °C to 460 °C range. In this range, $n_{Hall}$ is mainly determined by the fraction of P atoms that have been incorporated and electrically activated during the first thermal anneal at 350 °C. We attribute the continuous slow increase of $n_{Hall}$ from $4\times10^{13}$ cm$^{-2}$ at 210 °C to $7.6\times10^{13}$ at 460 °C to the higher electrical activation achieved because of the higher crystal quality of the encapsulation layer. At $T_g$ = 530 °C we observe a drastic ~ 2× increase of Hall density to a maximum of $1.3\times10^{14}$ cm$^{-2}$, corresponding to one P atom every ~5 Ge atoms in the dopant layer. The high electrical of activation of ~ 67%, calculated from $n_{Hall}/N_{SIMS}$, suggests that at 530 °C the pre-encapsulation anneal is crucial in dissociating the PH$_3$ and PH$_x$ molecules on the surface and efficiently incorporating P atoms into substitutional lattice sites where they become electrically active. This trend is confirmed for the sample encapsulated at 600 °C: all the donors that have not been desorbed from the surface during the pre-encapsulation anneal (~$1\times10^{14}$ cm$^{-2}$) are electrically active. More work is required to determine whether complete activation of carriers can be achieved at $T_g$ < 600 °C by optimizing the thermal budget of the first incorporation anneal step.

We shall now discuss the magnetotransport properties measured at $T$ = 4.2 K. Fig. 3 shows the zero magnetic field sheet resistance $\rho$ [panel (a)], the electron mobility $\mu$ (b), and the mean free path $l$ (c) as a function of $T_g$. We attribute the ~ 15× monotonic decrease in $\rho$ from 5.6 kΩ/□ to 0.372 kΩ/□ observed for $T_g$ increasing from 210 °C up to 530 °C to the higher carrier activation as a result of the higher crystal quality of the epitaxial encapsulation layer as shown in Fig. 1. At $T_g$ = 600 °C, there is a slight increase in $\rho$ to 0.448 k$\Omega$/□, mirroring the decrease in carriers caused by direct desorption form the pre-growth surface. The improvement in crystal quality with higher $T_g$ is also mirrored in the increase of the $\mu$. In particular we note a drastic increase in $\mu$ value from 65 cm$^2$/Vs for the 460°C sample to 128 cm$^2$/Vs for the 530 °C sample. The same trend is shown by the mean free path $l \propto \mu\sqrt{n_{Hall}}$, with a steep ~ 3× increase to ~ 12 nm at 530°C. The decrease in $l$ at the highest temperature (600 °C) is again attributed to the decrease in active carriers. In Fig 3(d) we report the magnetoconductivity $\Delta\sigma(B) = [\rho_{xx}(B)]^{-1} - [\rho_{xx}(0)]^{-1}$, where $\rho_{xx}(B)$ is the measured longitudinal resistivity as a function of $B$. For all the investigated samples we observe an inverted peak at zero field due to the weak localization from the coherent backscattering of electrons in time-reversed trajectories, highlighting the strong confinement of carriers in the dopant layer and the 2D nature of transport throughout the whole range of growth temperatures investigated. Along with the experimental curves, we report in Fig. 3(d) the theoretical fit to the Hikami model[18] for the magnetoconductivity of a disordered 2D system of non interacting electrons, obtained following the fitting procedure described in Ref. [1]. The obtained phase coherence length $l_\varphi$ vs. $T_g$ curve [Fig. 3(e)] shows the same trend observed for the mean free path, with a steep increase at $T_g$ = 530 °C to 180 nm caused by the higher crystal quality followed by a decrease at 600 °C due to the diminished $n_{Hall}$.

Comparing the 4.2 K electrical properties of Ge:P and Si:P $\delta$-layers at same carrier densities (~1.3×10$^{14}$ cm$^{-2}$), the values of peak mobility ~128 cm$^2$/Vs and phase coherence length ~180 nm obtained for the Ge:P system at $T_g$ = 530 °C exceed those reported for Si:P $\delta$-layers, which range from $\mu$ ~34 cm$^2$/Vs and $l_\varphi$ ~59 nm in Ref [19] to $\mu$ ~120 cm$^2$/Vs, $l_\varphi$ ~90 nm in Ref [8] depending on the study considered. The differences between Ge:P and Si:P $\delta$-layers at the same densities can be understood in terms of quality of the encapsulation layer. In Ge, the step-flow growth regime occurs at lower temperatures (500-600 °C) compared to silicon. This has the benefits of both embedding the P dopants in a higher quality, crystalline matrix whilst also mitigating dopant segregation. In contrast, Si must be grown at higher temperatures (650-800 °C) to achieve the same crystal quality,[20] which starts to become close to the P desorption temperature of ~ 650 °C. To avoid P segregation Si:P $\delta$-layers are typically encapsulated at lower temperatures (~250 °C),[7-9,12,19] which leads to an increase in surface roughness associated with the island-growth mode. In the Ge system this problem is not so acute since crystalline growth occurs at lower temperatures and dopant diffusion is mediated solely via vacancies compared with the silicon system where dopant diffusion is mediated via interstitial or vacancy diffusion or a mixed behavior of these two mechanisms.[21] The enhanced crystal quality and reduced dopant migration in Ge therefore leads to 2D $\delta$-layers with improved mobility and phase coherence.

In summary, we have investigated the effect of encapsulation temperature on the structural and electrical properties of Ge:P $\delta$-layers. For STM fabrication of future atomic-scale Ge devices, an encapsulation temperature of 530 °C represents the best compromise between enhanced electrical properties and minimal dopant redistribution. The large coherence length of 180 nm at 4.2 K indicate that Ge:P $\delta$-layers may be well suited for investigated novel quantum devices based on phase coherence. In addition, the high quality of the Ge epitaxial encapsulation layer due to growth in a step-flow regime suggests the use of the Ge regrowth surface for vertical stacked three-dimensional devices based on multiple $\delta$-layers.

GS acknowledges fruitful discussions with W. R. Clarke and O. Warschkow and support from UNSW under the 2009 Early Career Research and Science Faculty Research Grant scheme. GC is thankful to UNSW for a Visiting Professor Fellowship. MYS acknowledges an Australian Government Federation Fellowship.

**Figure 1:**

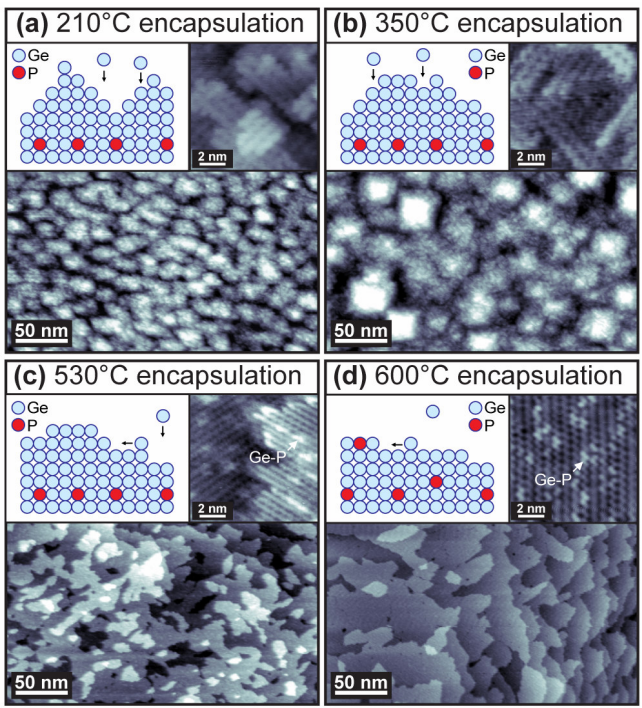

FIG. 1. (Color online) STM images of the surface topography of the δ-layers after encapsulation at different growth temperatures. Inset of each panel shows atomic resolution close-ups and schematic diagrams of the encapsulation process. The images were acquired at the following sample bias/tunneling current: -2.5 V/0.47 nA [(a), (b), inset in (b)], -2.1 V/0.23 nA [inset in (a)], -1.2 V/0.7 nA (c), -1 V/0.6 nA [inset in (c)], -2.3 V/0.25 nA (d), -1 V/0.7 nA [inset in (d)].

**Figure 2:**

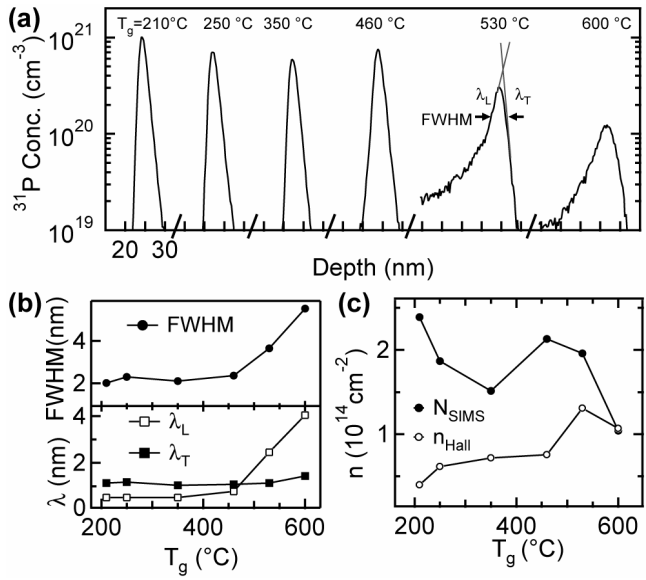

FIG. 2. (a) SIMS $^{31}$P depth profiles for δ-layers after encapsulation at different growth temperatures $T_g$. The traces are offset for clarity in the horizontal axis. (b) Peaks full width at half maximum (FWHM), leading edges ($\lambda_L$) and trailing edges ($\lambda_T$) as a function of $T_g$. (c) Total concentration of P atoms in the δ-layer ($N_{SIMS}$), obtained by integrating the SIMS depth profiles in (a), along with 4.2 K electrically-active carrier concentration $n_{Hall}$ from Hall effect as a function of $T_g$.

**Figure 3:**

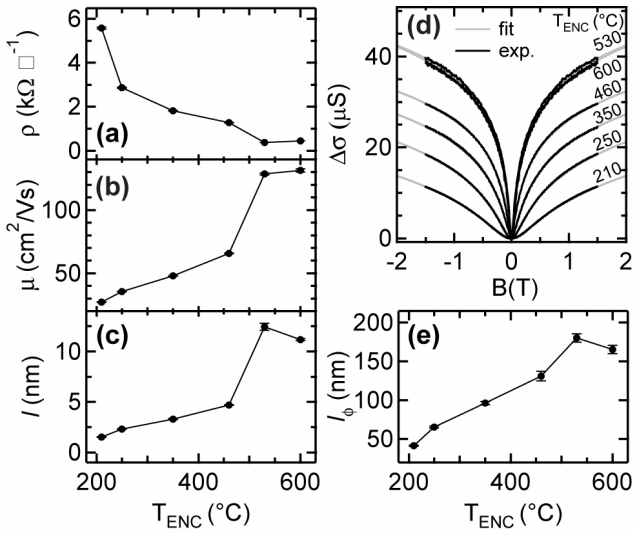

FIG. 3. 4.2 K (a) Sheet resistance $\rho$, (b) mobility $\mu$, (c) mean free path $l$ and (d) magnetoconductivities $\Delta\sigma$ as a function of growth temperature $T_g$. (e) Phase coherence length vs. $T_g$ from fitting the $\Delta\sigma$ curves in (d) to weak localization theory.